\documentclass[iop]{emulateapj}
\usepackage[colorlinks=true,linkcolor=red, citecolor=blue]{hyperref}
\usepackage{natbib}

\newcommand{\kms}{{\ensuremath{\text{ km s}^{-1}}}}

\newcommand{\FermiGO}{{\em Fermi Gamma Ray Space Telescope}}
\newcommand{\Fermi}{{\em Fermi }}
\newcommand{\upperRomannumeral}[1]{\uppercase\expandafter{\romannumeral#1}}

\newcommand{\distance}{{limit of \ensuremath{r_D > 6.5\text{ kpc}}}}
\newcommand{\percentgreater}{{\ensuremath{ > 40\%} }}
\newcommand{\percentgreatermod}{{\ensuremath{  40\%} }}

\newcommand{\fermimultiplier}{{\ensuremath{ \sim 2.1 } }}

\newcommand{\reddeningvalue}{{\ensuremath{ 1.16 \pm 0.12}}}
\newcommand{\reddeningJKvalue}{{\ensuremath{ 0.58 \pm 0.06} }}
\newcommand{\KExtinction}{{ 0.42} }

\slugcomment{Submitted for publication in the Astrophysical Journal Letters, \today}

\begin{document}
 \title{Distance and Reddening of the Enigmatic Gamma-ray-Detected Nova V1324 Sco}

\author{Thomas Finzell\altaffilmark{1}}
\author{Laura Chomiuk\altaffilmark{1}}
\author{Ulisse Munari\altaffilmark{2}} 
\author{Frederick M. Walter\altaffilmark{3}}

\altaffiltext{1}{Department of Physics and Astronomy, Michigan State Univeristy,  567 Wilson Road,  East Lansing, MI  48824-2320}
\altaffiltext{2}{INAF Astronomical Observatory of Padova, I-36012 Asiago (VI), Italy}
\altaffiltext{3}{Department of Physics and Astronomy, Stony Brook University, Stony Brook, NY 11794-3800}

\begin{abstract}
It has recently been discovered that some, if not all, classical novae emit GeV gamma-rays during outburst. Despite using an unreliable method to determine its distance, previous work showed that nova V1324 Sco was the most gamma-ray luminous of all gamma-ray-detected novae. We present here a different, more robust, method to determine the reddening and distance to V1324 Sco using high-resolution optical spectroscopy. Using two independent methods we derived a reddening of $E(B-V) = $ \reddeningvalue~and a distance \distance . This distance is \percentgreater greater than previously estimated, meaning that V1324 Sco has an even higher gamma-ray luminosity than previously calculated. We also use periodic modulations in the brightness, interpreted as the orbital period, in conjunction with pre-outburst photometric limits to show that a main-sequence companion is strongly favored.



\end{abstract}

\keywords{novae, cataclysmic variables, stars: distances}

\section{Introduction}
\label{sec:intro}
Classical nova are the result of a thermonuclear runaway taking place on the surface of a white dwarf and are fueled by matter accreted on to the white dwarf from a companion star. These outbursts give rise to an increase in luminosity and eject between $\sim 10^{-3} - 10^{-7} M_{\odot}$ of material at velocities $\gtrsim 10^3$ \kms~\citep{1986ApJ...310..222P, 1978ARA&A..16..171G, 2012BASI...40..185S}. 

Nova outbursts have also been detected in the GeV gamma-rays regime with \FermiGO\ (see e.g.~\citealt{2010ATel.2487....1C, 2012ATel.4224....1C, 2012ATel.4284....1C, 2013ATel.5302....1H, 2015ATel.7315....1C}). The likely source of these gamma-ray photons is strong shocks, which can generate relativistic particles via the diffusive shock acceleration mechanism and the relativistic particles can then generate gamma-rays~\citep{2014MNRAS.442..713M,1978ApJ...221L..29B}. But the details of shocks and shock acceleration in novae still remains a poorly understood issue, despite the potential for insight into the nature of these high-energy events~\citep{2015MNRAS.450.2739M}. 

The first nova to be detected by \Fermi was V407 Cyg, and it received considerable attention~\citep{2010Sci...329..817A, 2012ApJ...761..173C, 2012MNRAS.419.2329O, 2012ApJ...748...43N}. Given that V407 Cyg has a Mira giant secondary with a strong wind (a member of the symbiotic class of systems), a model to explain the gamma-rays was proposed wherein a shock was generated as the nova ejecta interacted with the dense ambient medium surrounding the red-giant companion.

This model, however, failed to explain V1324 Sco, V959 Mon, V339 Del~\citep{2014Sci...345..554A}, V1369 Cen~\citep{2013ATel.5649....1C}, and Nova Sgr 2015b~\citep{2015ATel.7315....1C}, all of which are novae that were detected by \Fermi but lack a detectable red-giant companion (hereafter referred to as non-symbiotic gamma-ray-detected novae which, for the sake of brevity, we will refer to as NGDN). This non-detection of a red-giant companion implies that these novae have main-sequence companions with low-density circumstellar material. We note that, while it is possible for these novae to have high density circumstellar material despite not having a red-giant companion, no evidence has yet been found for dusty circumstellar material around cataclysmic variables~\citep{2013AJ....145...19H}. The primary analysis for the NGDN was presented in~\citealt{2014Sci...345..554A}, which did not include V1369 Cen or Nova Sgr 2015b. Therefore the discussion that follows will only focus on the three non-symbiotic gamma-ray-detected nova discussed in~\citealt{2014Sci...345..554A}.

While no direct explanation was proposed by \citealt{2014Sci...345..554A} to explain the gamma-ray production in V1324 Sco, V959 Mon, and V339 Del, they did suggest that these novae were similar enough to one another that they represented a homogeneous population. Their evidence for a homogeneous population was that all three had similar gamma-ray spectra and light curves and were otherwise unremarkable novae. \citealt{2014Sci...345..554A} went on to suggest that it is only the close proximity of the NGDN that made them detectable by \emph{Fermi}, and that other novae would be detected if only they were closer. However, as we show in this paper, the NGDN are not actually homogeneous, as one of them clearly stands out from the rest.

V1324 Sco is located in one of the fields continually observed by the Microlensing Observations in Astrophysics (MOA) group, and its outburst was initially detected in 2012 May as part of their high-cadence $I$-band photometry~\citep{2012ATel.4157....1W}. The initial detection showed a slow monotonic rise in brightness between May 14 - May 31, followed by a very large ($\Delta I \sim 6$ mag) increase in brightness between June 1 - June 3~\citep{2012ATel.4157....1W}. Approximately two weeks after its optical brightening (2012 June 1, which we take to be the primary nova event) it was detected by the \Fermi collaboration as a new transient source~\citep{2012ATel.4284....1C}, with gamma-ray emission lasting a further $\sim$2 weeks~\citep{2014Sci...345..554A}. There are several things that make V1324 Sco unusual when compared to the other NGDN.

To begin with, V1324 Sco is the most gamma-ray luminous, being at least $\sim 2$ times greater than the other gamma-ray-detected novae (including V407 Cyg). The estimate for the gamma-ray luminosity, done by~\citealt{2014Sci...345..554A}, was based on distance estimations derived using the Maximum Magnitude Rate of Decline (MMRD) method, a technique that has been the subject of serious critiques in recent publications (see e.g.~\citep{2011ApJ...735...94K, 2012ApJ...752..133C}. The MMRD derived distance used in~\citealt{2014Sci...345..554A} was $\sim 4.5$ kpc, derived using the relation of~\citealt{1995ApJ...452..704D}. It should be noted that~\citealt{2014Sci...345..554A} do emphasize the uncertainty associated with this value, explicitly stating that a possible dust event could compromise their determination of the rate of decline in luminosity, which would alter their derived distance. In this paper we find a distance limit \percentgreatermod greater than that, meaning that the gamma-ray luminosity for V1324 Sco is \emph{even greater} than initially calculated.


Another unusual aspect of V1324 Sco is that it was never detected as an X-ray source. Along with diffuse shock acceleration, shocks also generate a tremendous amount of heat, raising the ejecta to $\sim$ keV temperatures~\citep{2015MNRAS.450.2739M, 2014MNRAS.442..713M}. This hot shocked gas can be observed by looking for hard X-rays~\citep{2001ApJ...551.1024M, 2008ApJ...677.1248M}. A stronger shock should generate more heat, which implies that the gamma-ray luminosity and X-ray luminosity should correlate with one another. However, despite the extreme gamma-ray luminosity of V1324 Sco, it was never detected as an X-ray source~\citep{2014MNRAS.442..713M}.

V1324 Sco clearly stands out among the NGDN, and this paper is focused on placing better constraints on this unique nova by finding values for the reddening and a distance limit. We present the results from our analysis of both spectroscopic and near-IR photometric data of V1324 Sco to find a distance limit, reddening, and constrain the luminosity class of the companion star. In Section~\ref{sec:ObsAndReduc} we present the details of the data observations and reduction. In Section~\ref{sec:Reddening} we detail the methods used to derive a value for the reddening, and in Section~\ref{sec:Distance} we provide similar details for our distance limit. In Section~\ref{sec:prog} we discuss the constraints we can place on the progenitor system and in Section~\ref{sec:concl} we conclude by deriving a new lower limit on the gamma-ray luminosity using our new distance limit.

\section{Observations}
\label{sec:ObsAndReduc}
To determine the reddening we analyzed two high-resolution spectroscopic observations of V1324 Sco; a very early time Very Large Telescope (VLT) observation, and a late time observation using the Magellan telescope. The VLT observation was taken from the archive\footnote{Based on observations made with ESO Telescopes at the Paranal Observatory under programme ID 089.B-0047, PI: Feltzing}, with the original observations reported by \citet{2012ATel.4157....1W}. The Magellan observations were obtained by the authors.

\subsection{VLT UVES} Spectra were taken on 2012 June 4.1 UT (+3.1 days after main outburst) by~\citealt{2012ATel.4157....1W} using the Ultraviolet and Visible Echelle Spectrograph (UVES) instrument on the VLT~\citep{2000SPIE.4008..534D}. The spectra were taken as part of the follow-up conducted by the MOA group. Copies of the data were made publicly available on the ESO archive and were obtained by the authors.

Observations were made in dichroic mode, with the blue arm centered at 4370 \AA\ (spanning $3600-4800$ \AA) and the red arm centered at 7600 \AA\ (covering $5600-9300$ \AA), taken using two CCDs that have a chip gap between $7550 - 7650$\AA. The slit width was $1.00^{\prime\prime}$ with an average seeing of $0.8^{\prime\prime}$, giving a resolution of  $R \approx 40,000$ for the blue arm and $40,000-50,000$ for the red arm. Four 1800 second integrations were taken with $1 \times 1$ binning, with each integration having an average signal-to-noise of $S/N \approx 30$ per pixel for the blue arm and $S/N \approx 110 - 175$ per pixel for the red.

The reduction was undertaken using the standard ESO Reflex data reduction pipeline, which includes flatfield correction, bias subtraction, cosmic-ray removal, spectral extraction, and wavelength calibration using comparison spectra of a ThAr lamp (see~\citealt{2014A&A...565A.113S} for details on the data reduction procedure). The four individual spectrum frames were combined using the IRAF routine \emph{ndcombine}, giving a final $S/N$ per pixel of $\approx 60$ in the blue and $S/N$ per pixel of $\approx 220 - 350$ in the red.

\subsection{Magellan MIKE} Further observations were made on 2012 July 16.1 UT  (+45.1 days after main outburst) using the MIKE instrument~\citep{2003SPIE.4841.1694B} on the 6.5 meter Magellan Clay telescope.

The MIKE instrument also has two arms, blue ($3350-5000$ \AA) and red ($4900-9500$ \AA).  Two 300 second integrations were taken with a $0.7^{\prime\prime}$ slit, giving a final resolution of $R \approx 40,000$ in the blue and $R \approx 30,000$ in the red. CCD binning was $2\times2$ for all integrations, giving an average $S/N$ per pixel of $\approx 30$ in the blue and $S/N$ per pixel of $\approx 170$ in the red.

The reduction for the MIKE spectra was done using the Carnegie Python tools (CarPy)\footnote{\url{http://code.obs.carnegiescience.edu/mike}}, which provides a simple pipeline data reduction procedure. Milky flats, Quartz flats, Twilight flats,ThAr comparison lamps and bias frames were all taken during the observation run and utilized in the data reduction pipeline.

\begin{center}
\begin{deluxetable}{cccr}
\tabletypesize{\scriptsize}
\tablewidth{2in}
\tablecaption{\label{tab:dibs}Best Fit Reddening Values For DIBs}
\tablehead{
\colhead{DIB $\lambda$ (\AA)}&
\colhead{EW  (m\AA)}&
\colhead{E($B-V$)}&
}
\startdata
5705.1& 99 $\pm$ 9 &  $1.01 \pm 0.11$ \\
5780.5	& 581 $\pm$ 40 &  $1.14 \pm 0.08$ \\
5797.1	& 155 $\pm$ 16 &  $0.86 \pm 0.09$ \\
6196.0	& 68 $\pm$ 2 &  $1.39 \pm 0.04$ \\
6613.6	& 200 $\pm$ 8 &  $0.94 \pm 0.04$ \\
\enddata
\vspace{-0.1in}
\end{deluxetable}
\end{center}
\begin{figure*}[t]
\includegraphics[width=\textwidth]{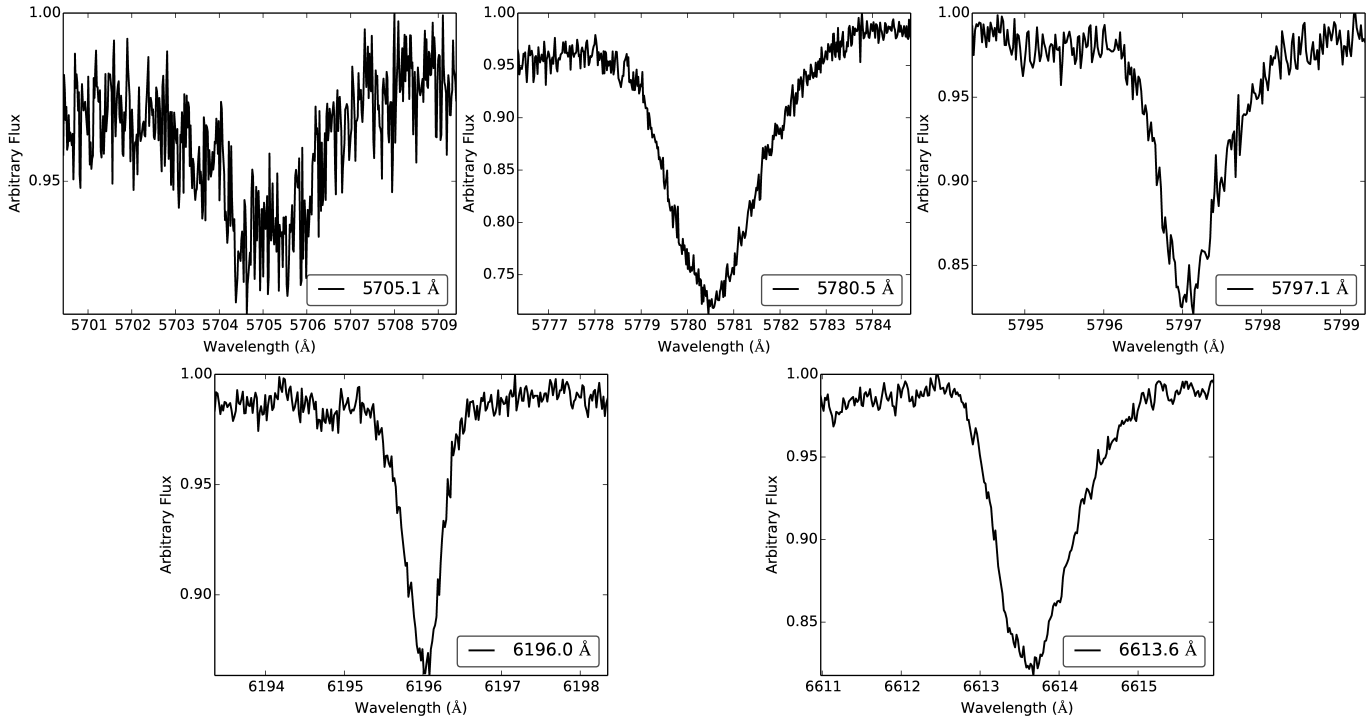}
\caption{\label{fig:DIBSGrid}
Five DIBs that were used to determine reddening. The largest contributor of uncertainty in the EW of the DIB features was determination of the continuum flux. This is why, in spite of the poor $S/N$, the 5705.1 \AA\ feature has a comparable uncertainty to the other features.
\vspace{0.1in}
}\end{figure*}

\begin{figure}[h]
\includegraphics[width=\columnwidth, trim=0mm 0mm 0mm 0mm,clip=true]{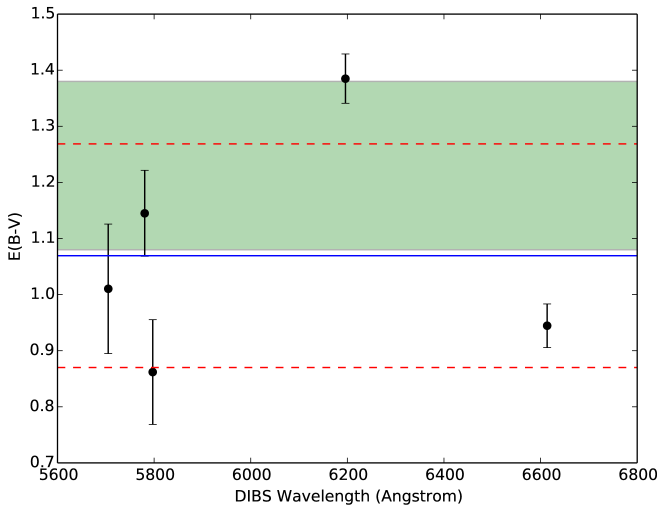}
\caption{\label{fig:BestFitDIBS}
The best fit $E(B-V) $ values for all five DIBs that were analyzed (see Table~\ref{tab:dibs} for the specific values). The reddening value for each DIB was determined using the best-fit values of~\cite{2011ApJ...727...33F}. The average reddening value is $E(B-V) = 1.07$ (the solid horizontal blue line) with an uncertainty of $\pm 0.20$ (the dashed horizontal red lines). The green band corresponds to the reddening value, $E(B-V) = 1.23 \pm 0.15$, derived using the EW of Na and K absorption features. The width of the band corresponds to the uncertainty in the derived value.
}\end{figure}
\begin{figure*}[t]
\includegraphics[width=\textwidth]{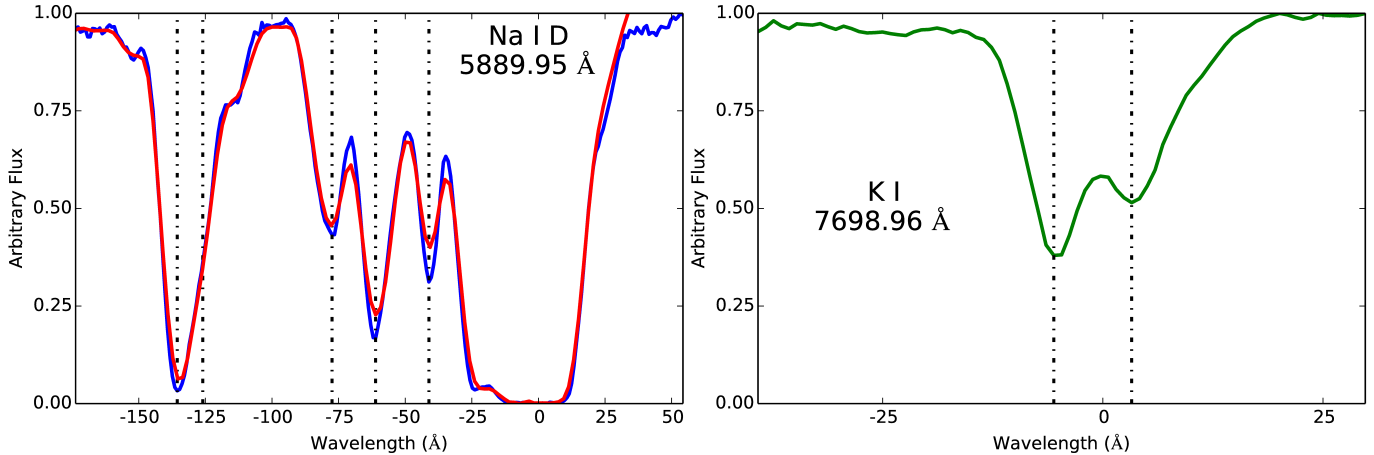}
\caption{\label{fig:NaAndKGrid}
Sodium D (5889.95~\AA ) and potassium (7698.96~\AA ) interstellar absorption features that were used to derive a value for the reddening. The specific absorption components from Table~\ref{tab:naandk} are marked as black dashed lines. We plot both the UVES and Magellan spectra for the sodium to show that these features are unchanging in time, meaning they are interstellar and not associated with the nova. Note that we only use the potassium absorption feature to calculate reddening when the sodium feature is saturated, which is why we are only showing the potassium in the region where the sodium is saturated.
}\end{figure*}

\section{Reddening}
\label{sec:Reddening}
\subsection{Reddening Measurement from DIBs}
We use the Equivalent Width (EW) of Diffuse Interstellar Bands (DIBs) to measure the reddening, a method that has already been used in  the context of novae (see e.g., \citealt{2011A&A...533L...8S, 2012ATel.4320....1M}). \cite{2011ApJ...727...33F} found an empirical relationship between the EW of eight strong optical DIBs and the reddening along a given sight line. In V1324 Sco, one of these eight DIB features (5487.7 \AA) had very low S/N (in both VLT and MIKE spectra) and was not used, and two features (6204.5 \AA\ and 6283.8 \AA) were discarded due to overlapping telluric features. The remaining five DIBs were used to find a value for $E(B-V)$. All of the EW values were measured using the VLT spectrum (see Figure~\ref{fig:DIBSGrid}) as it had a much higher $S/N$ in the red compared to the MIKE spectrum. The specific DIB EW values are given in Table~\ref{tab:dibs}, and the derived $E(B-V)$ as a function of wavelength is plotted in Figure~\ref{fig:BestFitDIBS}. All EW uncertainties were determined by iterative fitting solutions using the IRAF tool \emph{splot} to determine the spread of potential EW values. The primary uncertainty in determining the EW values was setting the continuum flux level. From our derived values for the reddening we took an error weighted average and found $E(B-V) = 1.07$ with a $1\sigma$ uncertainty on the error weighted average of $\pm 0.20$. The uncertainty for the individual $E(B-V)$  values derived for each DIB included both the uncertainty in the fit parameters from \cite{2011ApJ...727...33F} as well as the standard deviation in measured EW values. The final uncertainty in the error weighted average was dominated by the spread in derived $E(B-V)$ values for different DIB features.

\subsection{Reddening Measurement from Na and K Absorption Features}
We used a second, independent, method to determine the reddening. This method utilized the empirical relationship found by~\cite{1997A&A...318..269M}, which relates reddening to the total equivalent width of the Na I D absorption lines (at 5889.9 \AA\ and 5895.9\AA) and K I absorption line (at 7698.9 \AA). The material along the line of sight to an object will have some characteristic features due to interstellar absorption, and \cite{1997A&A...318..269M} calibrated two of these features to determine the amount of reddening each feature contributes. The sum of reddening from individual absorbing features gives the total reddening.

\cite{1997A&A...318..269M} found that the Na I D features are ideal for tracing reddening at low column densities, but saturates at high column densities. In these instances the K I feature, which does not saturate as easily, can be used to determine the reddening. In our case, we only needed to utilize the K I for the two Na I D features that were saturated (at $-7$ and $+3$ km/s).

To insure that the absorption features were the result of interstellar clouds, and not the nova itself, we compared the UVES and MIKE spectra. If the features were from the nova, we would expect them to change over time; as can be seen in the Na I D plot, which shows both the UVES and MIKE spectra, the features remained constant. We did not plot the MIKE spectra for the K I features, as the S/N was very low.

To avoid potential contamination of the lines with telluric features we used an archived telluric divider to remove telluric features from the UVES spectrum. The EW, derived reddening, and velocity of the absorption features that we used to find the total reddening are given in Table~\ref{tab:naandk} and Figure~\ref{fig:NaAndKGrid}. Although the uncertainty resulting from sum of the measured quantities is $\pm 0.09$, the intrinsic scatter that~\cite{1997A&A...318..269M} found in their calibration was $0.15$ for large reddening values ($E(B-V) \geq 0.4$). As a result, we take $0.15$ to be the uncertainty.

\begin{deluxetable}{cccc}
\tabletypesize{\scriptsize}
\tablecolumns{4} 
\tablewidth{0pt}
\tablecaption{\label{tab:naandk}Reddening Values from Na I D \& K I Absorption Lines}
\tablehead{
\colhead{Absorption} \vspace{-0.2cm}  & & & \colhead{LSR Radial} \\ \vspace{-0.2cm} 
& \colhead{EW  (m\AA)}&
\colhead{E($B-V$)}&  \\
\colhead{Wavelength (\AA)} & & & \colhead{Velocity (km s$^{-1}$)} 
}
\startdata
5886.94 & 246 $\pm$ 12 &  0.08 $\pm$ 0.01 &   $-136$ \\
5887.15 &  127 $\pm$ 6 &  0.04 $\pm$ 0.01 &  $-126$ \\
5888.06 & 148 $\pm$ 7 &  0.05 $\pm$ 0.01 &   $-79$ \\
5888.42 & 234 $\pm$ 12 &  0.08 $\pm$ 0.01 &   $-61$ \\
5888.82 & 171$\pm$ 9 &   0.05 $\pm$ 0.01 &   $-41$ \\
7698.37 & 103 $\pm$ 5 &  0.39 $\pm$ 0.02  &  $-7$ \\
7698.62 & 140 $\pm$ 7 &  0.54 $\pm$ 0.02 &   $3$ \\
 \hline \vspace{-0.25cm} \\ \vspace{-0.15cm}
 Total & & 1.23 $\pm$ 0.09 & \\ 
\enddata
\vspace{-0.1in}
\end{deluxetable}

The reddening value found using this technique, $E(B-V) = 1.23 \pm 0.15$, is consistent (within margin of uncertainty) with the reddening derived using the DIB features. 

We combine these two independent reddening measurements by taking an error weighted average of the two, which yields our final reddening value of $E(B-V) = $\reddeningvalue .

\section{Distance}
\label{sec:Distance}
In order to derive distance we used the reddening value found in Section~\ref{sec:Reddening} in conjunction with a 3D Galactic center reddening map from~\cite{2014A&A...566A.120S}, found using data from the \emph{Vista Variables in the Via Lactae} (\emph{VVV}) survey. The \emph{VVV} survey is a ESO large program using the 4-meter VISTA telescope to take near-IR photometry ($0.9-2.5 \mu$m) of 520 square degrees towards the Milky Way Bulge to characterize variable sources. The reddening map was one of the byproducts of the exquisite \emph{VVV} photometric dataset. The 3D map gives $E(J-K)$ reddening values as a function of: Galactic longitude (in intervals of $0.1^{\circ}$), Galactic latitude (in intervals of $0.1^{\circ}$), and radial distance (in intervals of 0.5 kpc, extending out to 10 kpc).

To make use of the map we needed to transform our $E(B-V)$ value into an $E(J-K)$ value given in the reddening map. To do this we found a coefficient, $\gamma$, such that $E(J-K) = \gamma E(B-V)$. The value of $\gamma$ was determined using the \emph{VVV} survey color transforms from Table 1 in~\cite{2012A&A...537A.107S}. These color transforms give the relative extinction for the \emph{VVV} filter system in terms of $E(B-V)$, assuming a standard extinction law of~\cite{1989ApJ...345..245C}. From these correction values we found a $\gamma$ value of 0.502, which gives $E(J-K) = $ \reddeningJKvalue for our reddening value of $E(B-V) = $\reddeningvalue .

\begin{figure}
\includegraphics[width=\columnwidth]{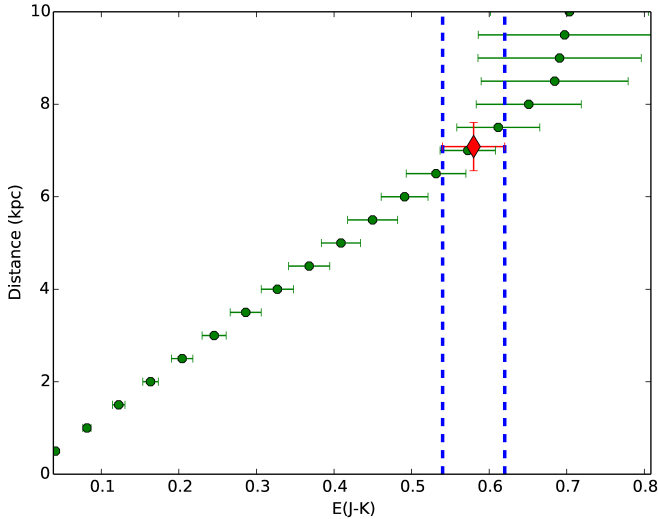}
\caption{\label{fig:ReddeningDistance}
The green data points give the (average) $E(J-K)$ reddening values as a function of distance at the position of V1324 Sco, taken from the 3D reddening map of \cite{2014A&A...566A.120S}. The red diamond indicates the derived reddening. The blue dashed lines show the $1\sigma$ extent of the derived reddening; it is clear that this value is consistent with all reddening values past 6.5 kpc. 
}\end{figure}

We used the average of the four points in the reddening map closest to the coordinates of V1324 Sco (RA $=$ 17:50:53.90, Dec $=$ $-$32:37:20.5) for our analysis. Figure~\ref{fig:ReddeningDistance} shows $E(J-K)$ versus radial distance $r_D$ for the reddening map values, as well as our derived value for V1324 Sco's $E(J-K)$ reddening. 

The dashed blue line in Figure~\ref{fig:ReddeningDistance} shows the $1\sigma$ maximum for our derived $E(J-K)$ reddening ($0.62$); it is clear that this value is consistent with any distance $> 6.5$ kpc. Because of this degeneracy in $E(J-K)$ reddening we can only place a lower limit of $6.5$ kpc on the distance to V1324 Sco.

\section{Progenitor}
\label{sec:prog}
If V1324 Sco had a giant companion like V407 Cyg then the wind from the companion could explain its exceptionally high gamma-ray luminosity. To determine if this is a feasible explanation we need to determine if the progenitor system to V1324 Sco has a giant secondary star. The strongest photometric constraints on the secondary star comes from the \textit{VVV} Survey~\citep{2010NewA...15..433M}, which obtained several epochs of near-IR photometry on the target field between July 2010 and September 2011. We searched the catalog for all stars with non-zero $K$-band magnitude in a 15'$\times$15' cutout centered on the coordinates of V1324 Sco ($N \approx 25,000$). To qualify as a star an object had to have a \emph{pStar} value $> 0.9$ (see \textit{VVV} documentation for a description of the \emph{pStar} variable). From this sample we found that 99\% of the sources had $m_{K} < 16.626$; given that V1324 Sco was not detected by the \textit{VVV} survey, we use this value as a limiting magnitude.

Table 1 in~\cite{2012A&A...537A.107S} gives the filter specific extinction for the \emph{VVV} $K$ band as $A_{K}/E(B-V) = 0.364$, and we can use our derived reddening value to obtain a $K$-band extinction of $\sim$ \KExtinction mags. Without a proper upper bound on the distance to V1324 Sco we cannot place an upper limit on the distance modulus. However we can say that, if V1324 Sco were at a distance of $\gtrsim 9.5$ kpc, then the distance modulus would be $\approx 15.3$, meaning that the absolute $K$-band limit would be bright enough for the companion to be a very faint giant ($M_K \geq 1.3$, corresponding to a spectral type G5 giant~\citep{2007AJ....134.2398C}). However, if V1324 Sco is at the galactic center ($\sim 8.5$ kpc), then it must have a dwarf companion.

While this limit isn't stringent enough to rule out a giant companion, by analyzing the period of V1324 Sco we show that the a main sequence companion is strongly favored.

\subsection{Constraints from the Orbital Period}
\cite{2012ApJ...746...61D} created a classification system for novae according to their orbital period which is a proxy for luminosity class of the secondary star. We utilize this classification scheme to help constrain the companion to V1324 Sco.

Measurements of the binary orbital period were found in the original photometry from the MOA group, who detected periodic modulations in the brightness of V1324 Sco~\citep{2012ATel.4157....1W}. These modulations were on the order of $\sim 0.1$ mag, with a period of $\sim 1.6$ hours. However, it is possible that there ellipsodial variations in the photometric light curve, which would give a secondary minimum and maximum. As a result, we consider both 1.6 and 3.2 hours for the period in our analysis.

Interestingly, if the period is 1.6 hours, V1324 Sco would be below the period gap, meaning that the angular momentum loss is driven by gravitational radiation~\citealt{2011ASPC..447....3K}. Only a handful of novae fall below the period gap~\citep{2012ATel.4157....1W}, and it has been postulated that these systems may have a different type of mass transfer mechanism that takes place in this period range~\citep{2010MNRAS.409..237U}.

\begin{figure}
\includegraphics[width=\columnwidth, trim=0mm 0mm 0mm 0mm,clip=true]{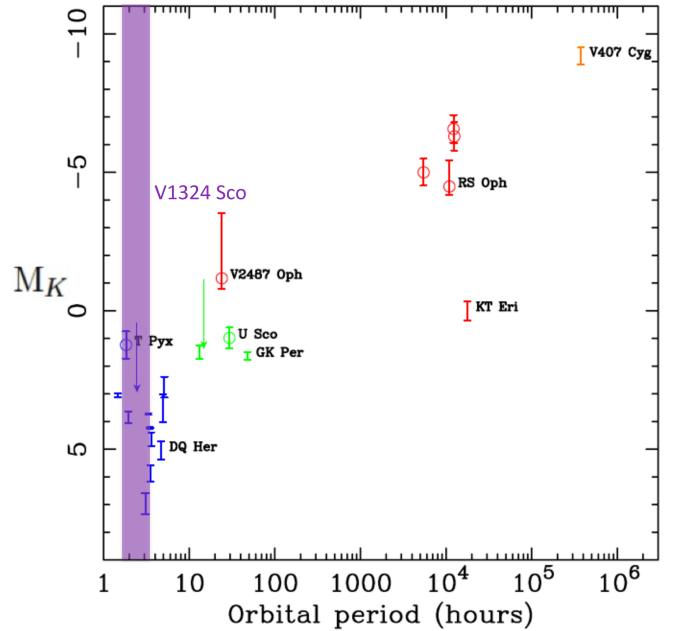}
\caption{\label{fig:Darnley}
Plot showing the distribution of novae as a function of absolute $K$-band magnitude and binary orbital period, taken from \citet{2012ApJ...746...61D}. V1324 Sco is marked as the purple transparent line, indicating that we know the period $\sim 1.6$ hours, but we cannot place a limit on the absolute $K$-band magnitude. The width of the purple line corresponds to the fact that the period could either be 1.6 or 3.2 hours. The colors on the plot correspond to the luminosity class of the secondary: blue is main sequence, green is sub-giant, and red is giant.
}\end{figure}

Using the value for the period within the framework of Darnley et al. (Figure~\ref{fig:Darnley}) we see that V1324 Sco (the purple dashed line) most likely falls into the region of blue points, indicting that it has a main sequence companion.

\section{Conclusions}
\label{sec:concl}
Here we present results of the analysis of optical spectroscopy and near-IR photometry of the nova V1324 Sco to determine the reddening and a limit on the distance. We derived a reddening value using two independent methods and found $E(B-V) = $ \reddeningvalue . Using this reddening value we found a distance lower limit of \distance ~to V1324 Sco, which stands in contrast to the values of~\cite{2014Sci...345..554A}, determined using the unreliable MMRD technique, of $4.5$ kpc. We also found that, while we cannot rule out a giant companion, the short period of V1324 Sco strongly favors a main sequence companion. V1324 Sco was already the most gamma-ray luminous nova when using the values from~\cite{2014Sci...345..554A}; using our new distance \distance  , we find that the gamma-ray luminosity is at least a factor of \fermimultiplier times greater than initially calculated by~\cite{2014Sci...345..554A}. We encourage deep follow-up photometry of V1324 Sco in order to place greater constraints on the companion.

\section*{Acknowledgments}

We are grateful for the of data from the ESO Public Survey program ID 179.B- 2002 taken with the VISTA telescope, data products from the Cambridge Astronomical Survey Unit, as well as data taken with the 6.5 meter Magellan Telescopes located at Las Campanas Observatory, Chile. This research has made use of the AstroBetter blog and wiki, and was funded in part by the \Fermi Guest Observer grant NNX14AQ36G  (L Chomiuk).

We are also especially grateful for the useful discussions with J. Linford, K. Mukai, T. Nelson, M. Rupen, J. Sokoloski, and R. Williams.

\bibliographystyle{apj}


\end{document}